\documentclass[aps,
12pt,
final,
notitlepage,
oneside,
onecolumn,
nobibnotes,nofootinbib,superscriptaddress,noshowpacs,centertags]
{revtex4-2}

\usepackage{amsmath,amssymb}
\RequirePackage{mathtext}
\RequirePackage[utf8]{inputenc}
\usepackage{hyperref}
\usepackage{caption2}
\usepackage{graphicx}

\begin{document}

\title{The project of a ground-based wide-angle\\EAS Cherenkov light imaging detector\\for PCR mass composition study\\in the 1--1000 PeV energy range}

\author{\firstname{D.\,V.}~\surname{Chernov}}
\email{chr@dec1.sinp.msu.ru}
\affiliation{Skobeltsyn Institute for Nuclear Physics, Lomonosov Moscow State University}
\author{\firstname{E.\,A.}~\surname{Bonvech}}
\email{bonvech@yandex.ru}
\affiliation{Skobeltsyn Institute for Nuclear Physics, Lomonosov Moscow State University}
\author{\firstname{O.\,V.}~\surname{Cherkesova}}
\affiliation{Skobeltsyn Institute for Nuclear Physics, Lomonosov Moscow State University}
\affiliation{Department of Space Research, Lomonosov Moscow State University}
\author{\firstname{E.\,L.}~\surname{Entina}}
\affiliation{Skobeltsyn Institute for Nuclear Physics, Lomonosov Moscow State University}
\author{\firstname{V.\,I.}~\surname{Galkin}}
\affiliation{Skobeltsyn Institute for Nuclear Physics, Lomonosov Moscow State University}
\affiliation{Faculty of Physics, Lomonosov Moscow State University}
\author{\firstname{V.\,A.}~\surname{Ivanov}}
\affiliation{Skobeltsyn Institute for Nuclear Physics, Lomonosov Moscow State University}
\affiliation{Faculty of Physics, Lomonosov Moscow State University}
\author{\firstname{T.\,A.}~\surname{Kolodkin}}
\affiliation{Skobeltsyn Institute for Nuclear Physics, Lomonosov Moscow State University}
\affiliation{Faculty of Physics, Lomonosov Moscow State University}
\author{\firstname{V.\,I.}~\surname{Osedlo}}
\affiliation{Skobeltsyn Institute for Nuclear Physics, Lomonosov Moscow State University}
\author{\firstname{N.\,O.}~\surname{Ovcharenko}}
\affiliation{Skobeltsyn Institute for Nuclear Physics, Lomonosov Moscow State University}
\affiliation{Faculty of Physics, Lomonosov Moscow State University}
\author{\firstname{D.\,A.}~\surname{Podgrudkov}}
\affiliation{Skobeltsyn Institute for Nuclear Physics, Lomonosov Moscow State University}
\affiliation{Faculty of Physics, Lomonosov Moscow State University}
\author{\firstname{T.\,M.}~\surname{Roganova}}
\affiliation{Skobeltsyn Institute for Nuclear Physics, Lomonosov Moscow State University}
\author{\firstname{M.\,D.}~\surname{Ziva}}
\affiliation{Skobeltsyn Institute for Nuclear Physics, Lomonosov Moscow State University}
\affiliation{Faculty of Computational Mathematics and Cybernetics, Lomonosov Moscow State University}


\begin{abstract}
This report presents a draft of a new detector designed to determine the chemical composition of primary cosmic rays based on the characteristics of the angular distribution of Cherenkov light from EAS. The installation, consisting of several such detectors, will be able to register individual EAS events in the energy range from 1 to 1000 PeV with high angular resolution of up to 0.2$^\circ$. The proposed detector’s distinctive feature is its simple design and wide viewing angle of above $\pm30^\circ$.
\end{abstract}

\maketitle

\section{Introduction}
Despite the large number of facilities operating in the `knee' energy range~\cite{GAMMA2007,Apel2012,Kampert2012,TAIGA2020}, the task of determining the chemical composition of primary cosmic rays (PCR) has not yet been fully solved~\cite{Schroeder2019,Schroeder2019b}. The methodological accuracy does not allow for individual classification of the recorded events. The most commonly used parameter, $X_\text{max}$, provides only an indication of the overall trend towards lighter or heavier composition~\cite{Yushkov2019,Omura2021}. However, this approach does not provide an unambiguous answer as to which of the p-He, CNO, or Fe nuclei groups is responsible for a particular change in the average $X_\text{max}$ value.

In an attempt to address this issue, a combination of methods involving Cherenkov facilities and particle detectors, such as Tunka, came in use to improve the separation of nuclei groups. Nevertheless, the use of the charged particle component of extensive air showers (EAS) inevitably leads to a significant dependence of the result interpretation on the initial model used for EAS simulation. At the same time, the Cherenkov light (CL) detection method has a significantly lower degree of model dependence, as it provides information about all stages of the cascade development, including the Cherenkov photons emitted by the primary particle. Although it is almost impossible to detect this light directly, but the angular distribution of CL can be used. Measuring the distribution of Cherenkov photons from different directions allows for a better identification of the primary particle type, as the information from various stages of the development of the cascade is obtained. This technique is particularly effective in astrophysical gamma-ray telescopes, where the use of the Hillas' parameters allows gamma quanta to stand out against the background of hadron components. 

However, there are technical and methodological challenges when studying hadrons. A technical challenge is the complexity and high cost associated with developing detectors for high-energy ranges. The technique requires the use of a large number of detectors, which must be wide-angled in order to achieve a sufficient geometric coverage factor. Additionally, these detectors must possess an angular resolution similar to that of gamma-ray telescopes in order to meet the requirements of the observation process. In order to address these challenges, we propose the development of a novel CL detector design.

\section{Detector design principle}
Relatively small detectors can be used to detect the hadron component of cosmic rays with energies in the range of 1--1000 PeV, as the photon density at a distance of 100 m from the shower axis is approximately 10 photons/cm$^2$ for a vertical 1~PeV proton shower. In order to reliably detect such an event, considering losses on optical elements, at least 1000 photons need to be collected. It is planned to utilize lenses with a diameter of about 22~cm and an area of $\sim$ 400~cm$^2$ for the prototype detector. However, a photodetector array with a large number of pixels is required in a classical telescope scheme to provide a wide field of view. Based on the fact that the occurrence of events with energies above 1~PeV is rare, and the likelihood of simultaneous occurrence of two such events from different directions is negligible, we propose the development of a detector that consists of a single, small matrix with up to 300--500 elements, which can be used to record light from several dozen lenses overlooking different directions at the same time.

\begin{table}[!ht]
\setcaptionmargin{0mm}
\onelinecaptionstrue
\captionstyle{flushleft}
\caption{A detector prototype paramenters.}
\label{tab:prop}
\bigskip
\begin{tabular}{|l|l|}
  \hline
   Field of view & $\pm28^\circ$ or $\pm36^\circ$ \\
  \hline
  Angular resolution& 0.2--0.4$^\circ$ \\
  \hline
  Focal length & 1.2--1.6~m \\
  \hline
  Detector diameter & 1.5--2~m \\
  \hline
  Pixel in the camera (virtual pixels for 37 lenses version) & 259 ($\sim$9\,000) or 427 ($\sim$15\,000) \\
  \hline
  Effective aperture & 320~cm$^2$\\  [1mm]
  \hline
\end{tabular}
\end{table}

\begin{figure}[t!]
\setcaptionmargin{5mm}
\onelinecaptionsfalse 
\includegraphics[width=1\linewidth]{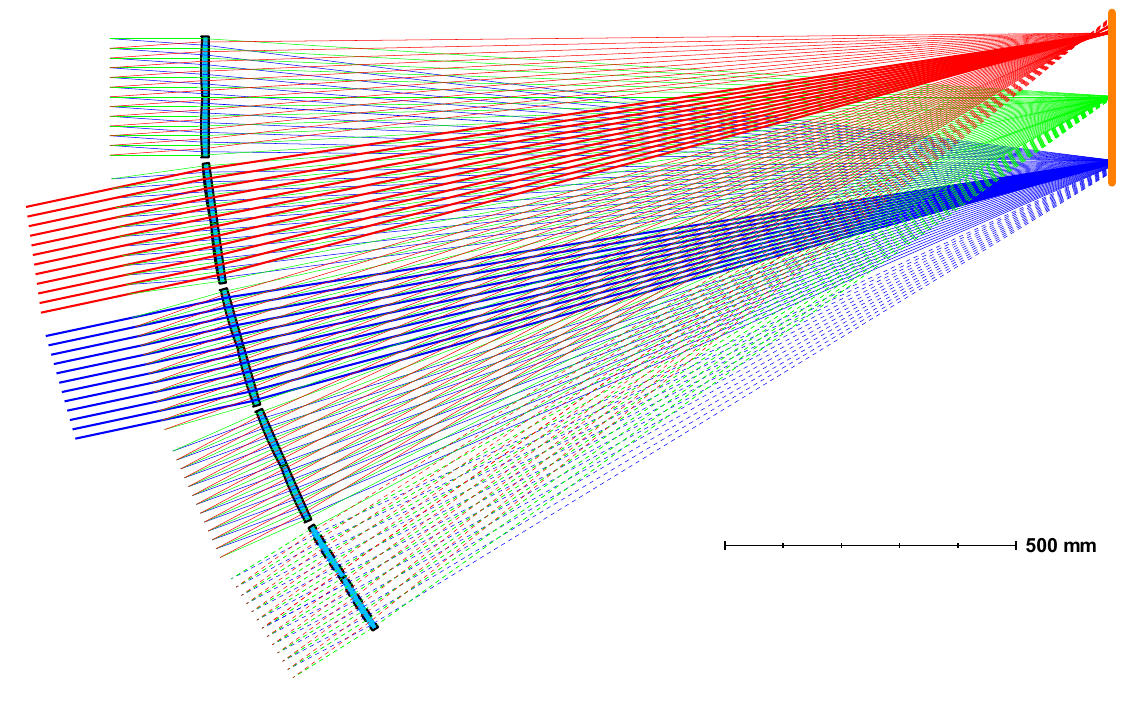}
\captionstyle{normal} \caption{The optical scheme of the proposed detector. Orange line indicates the detector's light sensitive camera. Light blue elements represent lenses. Each consecutive lens is tilted $8^\circ$ compared to its neighbours.  The coloured thin lines indicates the relative incidence angle on the lens: blue --- $-4^\circ$, green --- $0^\circ$, red --- $+4^\circ$. Bold red and blue lines show the parallel beams ($12^\circ$) that hit different lenses and are focused in different parts of the camera. A dashed line shows a possible additional lens (and even more can be added).}
\label{fig:optics}
\end{figure}

\begin{figure}[t!]
\setcaptionmargin{5mm}
\onelinecaptionsfalse 
\includegraphics{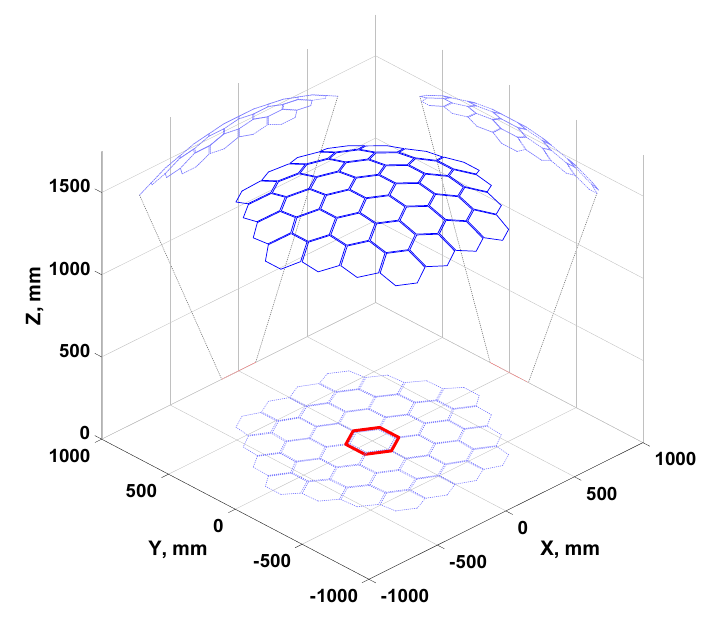}
\captionstyle{normal} \caption{General view of the detector. Solid lines show elements, dashed lines show side and top projections (given for clearer size comparison). Blue lines show lenses, red lines - detector camera. Dark gray dashed lines show side cover absent in central view.}
\label{fig:genview}
\end{figure}

Preliminary estimates have indicated the basic feasibility of creating such a detector. Fig.~\ref{fig:optics} show a preliminary version of the optical detector system, with an angular resolution of 0.24$^\circ$ and a focal length of 1550~mm, illustrating the path of rays for 4 lenses when a Cherenkov light hits at different angles. As the detector is symmetric (in see Fig.~\ref{fig:genview} for detector general view), the total field of view will be greater than 56$^\circ$. The upper right corner (orange rectangle) depicts a photodetector array with a diameter of about 285~mm. In total, 37 identical convex lenses with a curvature radius of 803~mm are mounted in the detector. Each lens focuses light from an some area within the full field of view onto the camera. To ensure the continuous sensitivity across the whole field of view the overlap of individual lenses fields of view is provided. In Fig.~\ref{fig:optics} bold red and blue lines demonstrate that parallel beams from adjacent lenses are collected on opposite edges of the camera. The number of lenses and, therefore, detector viewing angle can be further increased to 72$^\circ$ by adding another row of lenses (61 lenses in total). And by replacing each lens by a two lens block, another increase in detector viewing angle can be made, bringing total field of view almost to 1.8~sr. 

The camera consists of 61 optical modules. Optical module houses 7 silicon $6\times6$~mm silicon photomultipliers (SiPM) with individual spherical lens light collectors, arranged in a hexagonal pattern. A center-to-center distance between neighbouring SiPMs is approximately 12~mm. The change in size of the camera with fixed lens system changes the individual lens field of view and, since the lenses orientation is fixed, in higher or lower overlap. The change in pixel density changes resolution.

A similar device is also under development, with an angular resolution of 0.33$^\circ$, viewing angle of $\pm20^\circ$, and a focal length of 1600 mm. In all versions of the prototype, the same SiPM module will be employed.

The optical module is based on the SensL MicroFC-60035 SiPMs with $5.96\times5.96$~mm$^2$ area. A similar optical module of 7 SiPMs has been developed previously~\cite{Bonvech2019}. At the end of 2018, a first camera prototype (7 modules totalling 49 SiPMs) underwent testing, and since September 2019, this prototype (Small Imaging Telescope, SIT) has been successfully integrated into the TAIGA astrophysical complex~\cite{Chernov2020,Podgrudkov2021}.

\section{Sensitivity to primary particle type}

The analysis of the EAS CL angular distribution is commonly used in gamma-ray astronomy to filter events from primary gammas using so-called `Hillas parameters' as described in~\cite{Hillas1996}. However, in gamma-astronomy detectors mirrors has diameters ranging from several meters to tens of meters all in order to achieve a low energy threshold in the hundreds of MeV range and below. For energies around 1~PeV direct applicability of this method is a matter of a further careful study. However, for mass composition studies another set of methods was developed for relatively high altitude conditions of mountain station~\cite{Borisov2013,Galkin2017,Galkin2018,Bakhromzod2021}. Adaptation of these methods to the deeper levels is done in the SPHERE-3 project~\cite{Chernov2022,Chernov2024}. 

Preliminary, a proposed detector can divide registered events into several groups if some additional data will be available (see Sec.~\ref{sec:application}). 

\section{Energy threshold}
Since the useful signal is typically generated by only a single lens, while background signals are collected simultaneously from all detector lenses, it is necessary to determine the signal-to-background ratio. Assuming a typical night sky photon flux estimation of $2\!\cdot\!10^{12}$ photons~m$^{-2}$~sr$^{-1}$~s$^{-1}$, the photon flux through a 320~cm$^2$ aperture and 0.24$^\circ$ pixel ($1.38\!\cdot\!10^{-5}$~sr) times 37 lenses for 200 ns will be less than 8. Considering the SiPM's quantum efficiency to be approximately 25\%, background fluctuations will range from 2--3 photoelectrons.

However, a more accurate estimation can be obtained from analysis of the SIT telemetry data ~\cite{Chernov2020}. Using absolute calibration~\cite{Amineva2023}, the total camera current can be converted to a flux of $1.78\!\cdot\!10^9$ photoelectrons/s (after subtracting any additional current from SiPM crosstalk). This gives a value of $3.7\!\cdot\!10^7$ photoelectrons/s per pixel.This estimation is more accurate because it takes into account the spectral dependencies of the SiPM and the real spectral parameters of the night sky photon flux.

Considering the difference in effective aperture (SIT --- 367~cm$^2$, this detector --- $\sim320$~cm$^2$) and pixel size (SIT --- 1.5~$\Box^\circ$, this detector --- $\sim$0.05~$\Box^\circ$) the expected number of background photoelectrons per 200 ns is 0.2, which makes the background negligible with respect to signal reconstruction\footnote{It should be noted that for trigger system this background level in conjunction with the SiPM crosstalk poses a significant challenge. But there are ways to circumvent this issue. See details and some possible solutions in~\cite{Entina2024} for a more challenging situation of long signal collection times and higher background.}.

The same comparison with SIT data~\cite{Podgrudkov2021} allows us to estimate a preliminary energy threshold that is comparable with SIT (with almost the same aperture). As the direct CL from an EAS has a narrow distribution, in cases where we are not far from the shower axis, most of the photons will be collected within 4--5 pixels. The SIT detector has a lower energy threshold of 0.3~PeV (obtained through cross-correlation with HiSCORE). Therefore, this new detector will likely have an energy threshold close to 1~PeV.

However, a full Monte Carlo simulation may refine this estimation. Scaling the detector by a factor of 2 should also reduce the energy threshold to approximately 250--300 TeV.


\section{Detector design applicability}
\label{sec:application}

The proposed detector has certain limitations regarding its use. Due to the fact that each pixel of the detector simultaneously observes several different spots on the celestial sphere (one spot per lens), it is not technically possible to determine from which lens light was collected during an event. Therefore, it is not possible to reconstruct the arrival direction of the cosmic ray (CR) based on data from the camera of the detector alone, and consequently, it is impossible to reliably determine energy and particle type.

There are several ways to solve this problem. One option is to integrate the new detector with another extensive air shower (EAS) array and use external data on shower arrival direction. The arrival directions of EAS and CR are different, so external data may not be very accurate (within 5$^\circ$). In such a case, the external data on EAS arrival direction can be used for particle type estimation.

\begin{figure}[t!]
\setcaptionmargin{5mm}
\onelinecaptionsfalse 
\includegraphics[width=0.5\linewidth]{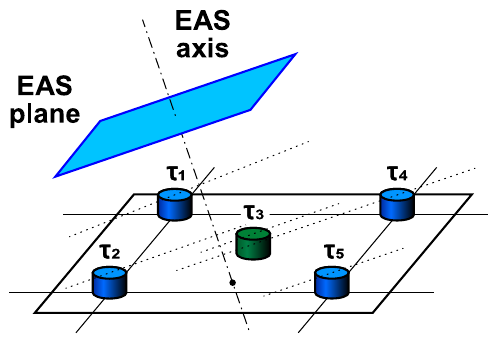}
\captionstyle{normal} \caption{Array of several detectors. Green one indicated the proposed detector. Blue ones can be of the same design or a simple wide-angle `single-pixel' detector with sole purpose to provide time information on shower passing. Alternatively blue detectors can be from a different independent array. Detector spacing is about 100--200~m.}
\label{fig:array}
\end{figure}

Another option is to equip the detector with additional detector units. These auxiliary units, located 100--200 meters from the main detector and consisting of a single pixel, would serve to reconstruct the arrival direction and, in the case of more complex designs, the location of the axis. The layout and operation of these units is shown in Fig.~\ref{fig:array}. The green unit represents a proposed detector, and the blue units represent auxiliary detectors.

This approach can be further expanded by using not a single additional detector, but a grid of them. For example, copies of the main detector can be used as auxiliary units (blue units in Fig.~\ref{fig:array}). This would provide data on the arrival direction of the shower (from photon arrival times), information on the location of the axis (using several methods), and additional information on the type of particle.

Simultaneous detection of the same EAS by multiple detectors would allow for more sophisticated data analysis techniques, as the angle between the arrival of the shower and CL arrival also depends on the particle type.


\section{Conclusion}
A design for a wide-angle ground base detector with high angular resolution for PCR mass composition study above 1~PeV energy is proposed. The optical parameters of the proposed detector meet the requirements for angular resolution. Evaluation of the influence of the background of the starry sky shows the possibility of registering EAS events from particles with an energy of 1~PeV.

\begin{acknowledgments}
Prototype tests were done using facilities of Crimean astrophysical observatory and of Institute of Astronomy of the Russian Academy of Sciences in Simeiz. 
\end{acknowledgments}


\begin{thebibliography}{99}

\bibitem{GAMMA2007}
A.P.~Garyaka, R.M.~Martirosov, S.V.~Ter-Antonyan, N.~Nikolskaya, Y.A.~Gallant, L.~Jones and J.~Procureur, Astropart. Phys., \textbf{28}(2), 169 (2007). https://doi.org/10.1016/j.astropartphys.2007.04.004

\bibitem{Kampert2012}
K.-H.~Kampert and M.~Unger, \textbf{35}(10), 660 (2012). https://doi.org/10.1016/j.astropartphys.2012.02.004

\bibitem{Apel2012}
W.D.~Apel, J.C.~Arteaga-Velazquez, K.~Bekk, M. Bertaina, J.~Bl\"{u}mer, H.~Bozdog, I.M.~Brancus, P.~Buchholz, E.~Cantoni, A.~Chiavassa, F.~Cossavella, K.~Daumiller, V.~de Souza, F.~Di~Pierro, P.~Doll, R.~Engel et al., Astropart. Phys. \textbf{36}(1), 183 (2012). https://doi.org/10.1016/j.astropartphys.2012.05.023

\bibitem{TAIGA2020}
N.~Budnev, I.~Astapov, P.~Bezyazeekov, E.~Bonvech, V.~Boreyko, A.~Borodin, M.~Br\"{u}ckner, A.~Bulan, D.~Chernov, D.~Chernykh, A.~Chiavassa, A.~Dyachok, O.~Fedorov, A.~Gafarov, A.~Garmash, V.~Grebenyuk et al., JINST \textbf{15}(09), 09031 (2020). http://doi.org/10.1088/1748-0221/15/09/c09031

\bibitem{Schroeder2019}
F.G.~Schr\"{o}der, T.~AbuZayyad, L.~Anchordoqui, K.~Andeen, X.~Bai, S.~BenZvi, D.~Bergman, A.~Coleman, H.~Dembinski, M.~DuVernois, T.~Gaisser, F.~Halzen, A.~Haungs, J.~Kelley, H.~Kolanoski, F.~McNally et al., BAAS, \textbf{51}(3), 131 (2019) https://baas.aas.org/pub/2020n3i131

\bibitem{Schroeder2019b}
F.G.~Schr\"{o}der, PoS(ICRC2019), \textbf{358}, 030 (2019) https://doi.org/10.22323/1.358.0030

\bibitem{Yushkov2019}
A.~Yushkov, J.~Bellido, J.~Belz, V.~de~Souza, W.~Hanlon, D.~Ikeda, P.~Sokolsky, Y.~Tsunesada, M.~Unger, EPJ Web of Conf. \textbf{210}, 01009 (2019). https://doi.org/10.1051/epjconf/201921001009

\bibitem{Omura2021}
Y.~Omura, R.~Tsuda, Y.~Tsunesada, D.R.~Bergman and J.~Krizmanic,  PoS(ICRC2021), \textbf{3395}, 329 (2021). https://doi.org/10.22323/1.395.0329

\bibitem{Bonvech2019}
E.~Bonvech, T.~Dzhatdoev, D.~Chernov, Mir.~Finger, Mih.~Finger, V.~Galkin, G.~Garipov, V.~Kozhin, D.~Podgrudkov, A.~Skurikhin, J. Phys. Conf. Ser. \textbf{1181}, 012025 (2019). https://doi.org/10.1088/1742-6596/1181/1/012025

\bibitem{Chernov2020}
D.~Chernov, I.~Astapov, P.~Bezyazeekov, E.~Bonvech, A.~Borodin, M.~Br\"uckner, N.~Budnev, D.~Chernukh, A.~Chiavassa, A.~Dyachok, O.~Fedorov, A.~Gafarov, A.~Garmash, V.~Grebenyuk, O.~Gress, T.~Gress et al., JINST \textbf{15}(09), 09062 (2020). https://doi.org/10.1088/1748-0221/15/09/c09062

\bibitem{Podgrudkov2021}
D.A.~Podgrudkov, E.A.~Bonvech, I.V.~Vaiman, D.V.~Chernov, I.I.~Astapov, P.A.~Bezyazeekov, M.~Blank, A.N.~Borodin, M.~Br\"uckner, N.M.~Budnev, A.V.~Bulan, A.~Vaidyanathan, R.~Wischnewski, P.A.~Volchugov, D.M.~Voronin, A.R.~Gafarov et al., Bull. Rus. Acad. Sci.: Phys. \textbf{85}(4), 408 (2021). https://doi.org/10.3103/s1062873821040286

\bibitem{Hillas1996}
A.M.~Hillas, Space Sci. Rev. \textbf{75}(1/2), 17 (1996). https://doi.org/10.1007/BF00195021

\bibitem{Borisov2013}
A.S.~Borisov, V.I.~Galkin, M.I.~Ilolov, R.A.~Mukhamedshin, H.H.~Muminov, V.S.~Puchkov, O.~Saavedra, Proc. 33\textsuperscript{rd} ICRC, Rio de Janeiro, 0953 (2013).

\bibitem{Galkin2017}
V.I.~Galkin, A.S.~Borisov, R.~Bakhromzod, V.V.~Batraev, S.~Latipova, A.~Muqumov, EPJ Web of Conf. \textbf{145}, 15004 (2017). https://doi.org/10.1051/epjconf/201714515004

\bibitem{Galkin2018} 
V.I.~Galkin, A.S.~Borisov, R.~Bakhromzod, V.V.~Batraev, S.~Latipova, A.~Muqumov, Moscow Univ. Phys. Bull. \textbf{73}(2), 179 (2018). http://doi.org/10.3103/S0027134918020078

\bibitem{Bakhromzod2021}
R.~Bakhromzod, V.I.~Galkin, NIM A \textbf{1018}, 165842 (2021). http://doi.org/10.1016/j.nima.2021.165842

\bibitem{Chernov2022}
D.V.~Chernov, C.G.~Azra, E.A.~Bonvech, V.I.~Galkin, V.A.~Ivanov, V.S.~Latypova, D.A.~Podgrudkov, T.M.~Roganova, Phys. At. Nucl. \textbf{85}(6), 641 (2022). https://doi.org/10.1134/S1063778822060059

\bibitem{Chernov2024}
D.V.~Chernov, C.J.~Azra, E.A.~Bonvech, O.V.~Cherkesova, E.L.~Entina, V.I.~Galkin, V.A.~Ivanov, T.A.~Kolodkin, N.O.~Ovcharenko, D.A.~Podgrudkov, T.M.~Roganova, M.D.~Ziva, Phys. At. Nucl. \textbf{87}(S2), S319 (2024). https://doi.org/10.1134/S1063778824700959

\bibitem{Amineva2023}
A.A.~Amineva, A.V.~Pantiukhin, D.A.~Podgrudkov, Memoirs Fac. Phys. \textbf{2023}(4), 2341602

\bibitem{Entina2024}
E.L.~Entina, D.A.~Podgrudkov, C.G.~Azra, E.A.~Bonvech, O.V.~Cherkesova, D.V.~Chernov,
V.I.~Galkin, V.A.~Ivanov, T.A.~Kolodkin, N.O.~Ovcharenko, T.M.~Roganova, M.D.~Ziva, Moscow Univ. Phys. Bull. \textbf{79}(Suppl. 2), S676 (2024). http://doi.org/10.3103/S0027134924702126

\end{thebibliography}
\end{document}